\documentclass{cup-hpl}

\usepackage[printonlyused]{acronym} 
\usepackage{makecell}
\usepackage{array}
\usepackage{tabularx}

\begin{document}


\shorttitle{Machine Learning in LPA}                                   
\shortauthor{J. Lin et al.}

\title{Applications of object detection networks at high-power laser systems and experiments}

\author[1]{Jinpu Lin}
\author[1]{Florian Haberstroh}
\author[1]{Stefan Karsch}
\author[1]{Andreas Döpp}

\address[1]{Ludwig-Maximilians-Universit\"at M\"unchen, Am Coulombwall 1, 85748 Garching, Germany}

\begin{abstract}
The recent advent of deep artificial neural networks has resulted in a dramatic increase in performance for object classification and detection. While pre-trained with everyday objects, we find that a state-of-the-art object detection architecture can very efficiently be fine-tuned to work on a variety of object detection tasks in a high-power laser laboratory. In this manuscript, three exemplary applications are presented. We show that the plasma waves in a laser-plasma accelerator can be detected and located on the optical shadowgrams. The plasma wavelength and plasma density are estimated accordingly. Furthermore, we present the detection of all the peaks in an electron energy spectrum of the accelerated electron beam, and the beam charge of each peak is estimated accordingly. Lastly, we demonstrate the detection of optical damage in a high-power laser system. The reliability of the object detector is demonstrated over one thousand laser shots in each application. Our study shows that deep object detection networks are suitable to assist online and offline experiment analysis, even with small training sets. We believe that the presented methodology is adaptable yet robust, and we encourage further applications in high-power laser facilities regarding the control and diagnostic tools, especially for those involving image data.
\end{abstract}

\keywords{Laser-plasma accelerators, machine learning, object detection, optical diagnostics, image processing}

\maketitle

\section{Introduction}
High power laser systems with power reaching the petawatt-level and repetition rate at a fraction of a hertz have emerged worldwide in the past few years\cite{sung20174, cala, nees2020zeus, zhang2020laser, borneis2021design}.
With the fast development of high-repetition-rate operation capabilities in plasma targetry, high-power laser-plasma experiments can employ statistical methods that require a large number of shots. Studies for real-time optimization using evolutionary algorithms have been reported in recent years\cite{he2015coherent, streeter2018temporal, lin2019adaptive, smith2020optimizing, englesbe2021optimization}. As the size of data to process has continued to increase, more advanced machine learning models have attracted increasing attention. By constructing predictive models, machine learning methods are employed to model the nonlinear, high-dimensional processes in high-power laser experiments. Various methods, including neural networks, Bayesian inference, and decision trees have been introduced for optimization tasks and physics interpretation\cite{humbird2018deep, gonoskov2019employing, humbird2019transfer, hsu2020analysis, shalloo2020automation, lin2021beyond}. Meanwhile, as the measurement and diagnostic tools evolve, digital imaging is playing increasingly important roles in experiments and with it, machine learning methods to process image data.

In the case of a laser-plasma accelerator image-based diagnostics can take a variety of forms, starting from the optical elements in the high-power laser facility, over shadowgraphy and interferometry of plasma dynamics, to scintillator signals generated by energetic electron or X-ray beams from the accelerator. In particular, the evolving structure of a plasma accelerator is challenging to visualize because of its microscopic size ($\sim10^{-5}$ m) and its high velocity (approaching the speed of light). With the latest techniques such as few-cycle shadowgraphy, taking snapshots of the plasma wake structure is enabled in femtosecond resolution over a range of picoseconds \cite{savert2015direct, gilljohann2019direct, ding2020nonlinear}. The latest generation of laboratory diagnostics for the plasma structures is reviewed by Downer \textit{et al.} \cite{downer2018diagnostics}.

In this manuscript, we demonstrate exemplary applications of an object detection network in the diagnostics in a high-power laser laboratory. We apply the object detector to few-cycle shadowgraphy of plasma waves, to an electron energy spectrometer, and to detect optical damages in a high power laser beamline. The results show that object detection enables possibilities in diagnostics and data analysis that have not yet been achieved using conventional methods. Moreover, due to the fast inference speed of the object detector, it paves the road towards real-time demonstration of such diagnostics during experiments.

\section{Object detection algorithms}
\begin{figure*}[ht]
\centering
\includegraphics[width=1\linewidth]{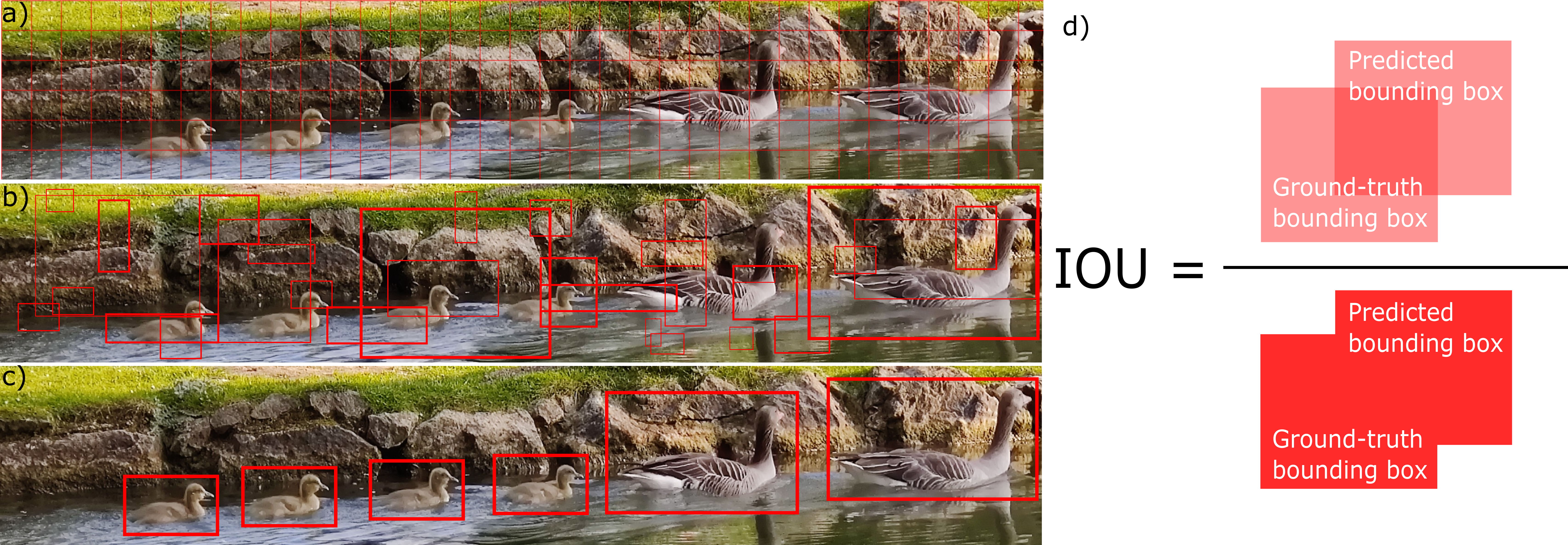}
\caption{Step-wise illustration of the object detection method. The example image presents ducks creating and surfing on wakefields. (a) Split the image into $S\times S$ grids; (b) Predict bounding boxes and confidences for each class; (c) final detected objects with confidences; (d) Bounding box predicted by the object detector vs. the ground-truth bounding box labeled manually. IoU is defined as their area of intersection divided by their area of union, while an ideal object detector would have IoU=1.
}
\label{intro}
\end{figure*}

Since the development of convolutional neural networks (CNNs), computer vision has drawn attention from across various disciplines\cite{gu2018recent, yamashita2018convolutional, li2021survey}. 
As a huge breakthrough in image recognition, a CNN allows categorizing images into certain classes. When a CNN classifies an input image, it learns a model that detects the specific patterns on that image. A pattern is detected by a "filter" matrix, which has a pre-defined size relatively smaller than the size of the input image. It then takes the dot product of the filter matrix with a sub-matrix of the input image (in pixel values) that has the same size. The filter is "convolved" with the input image as it slides across the entire input image matrix for all sub-matrix of its size, resulting in an output matrix of cross products. Intuitively, a filter in a CNN is analogous to a neuron in a regular feed-forward neural network, and several filter matrices form into a convolutional layer. A complex CNN can have multiple convolutional layers, and the final output matrix is compared to the input image to adjust the values of the filter matrices. This process is repeated over and over until the output matrix is close enough to the input.

An extension to classification tasks in computer vision is object detection. Unlike classification tasks such as image recognition, which assign one single label to the image, object detection aims to identify all the objects of interest in an image, classify each object and assign a label to it, and then locate them by drawing a bounding box around each object. For images with fixed number of objects, the objects can be found using a standard CNN followed by a fully-connected output network layer with a pre-specified length. However, the task becomes much more challenging when the number of interesting objects is not fixed in an image, leading to a varying length of the output layer of the neural network. This happens to be the case for most applications in high-power laser experiments, especially when scanning parameter spaces across various laser and plasma conditions.

Theoretically, the problem can be solved by splitting the image into many regions of interest and coupling a CNN to each region. However, the number of regions could be significant and easily exceed the computational limit. To make it computationally efficient, there are two families of methods to locate and label objects without determining the number of objects in advance. The region-based convolutional neural network\cite{girshick2015region} (R-CNN) and its later iterations (Faster R-CNN, Mask R-CNN) use a selective algorithm to propose a reasonable amount of regions that may contain bounding boxes. It then applies a CNN to extract features from each candidate region and classify the feature into the known classes using a linear classifier. While R-CNN is very accurate in locating the objects, its computational cost can be heavy.

The "You Only Look Once" (YOLO) family of algorithms \cite{redmon2016you} take a different approach, where a simplified methodology is illustrated in Fig. \ref{intro}a-c. YOLO splits the image into a pre-determined number of grids, and defines multiple bounding boxes for each grid. Unlike R-CNN which applies a network to each region, YOLO applies a single neural network to the full image. The network then predicts a probability for each class for each bounding box. Post-processing is performed to select the best bounding boxes based on the probability and the overlapping conditions regarding their neighbouring boxes. The biggest advantage of YOLO, as its name suggests, is that it makes predictions with a single network evaluation instead of thousands in other methods like R-CNN. Therefore, YOLO can be two or three orders of magnitude faster than R-CNN, making it possible for real-time object detection tasks. This is of particular interest to the community of high-power laser experiments, especially with the development of high-repetition-rate capabilities. However, it has to be pointed out that YOLO's superiority in efficiency comes at the cost of prediction accuracy, such as in locating the bounding boxes.

As a supervised learning task, validation is needed after training an object detection model. The commonly-used evaluation metric in object detection is the Intersection Over Union (IoU). To evaluate the model accuracy on a predicted bounding box, we manually label a ground-truth bounding box and the IoU calculates the area of the intersection as well as the area of the union, see Fig. \ref{intro}d. The ratio of these two areas is defined as the IoU value between 0 and 1, where 0 means no intersection and 1 means completely overlapping. For a set of images, the performance of the object detection model is evaluated using the mean average precision (mAP), which is obtained by averaging over different IoU thresholds on each bounding box on each image. The box confidence score C is then defined as
\begin{equation}
\label{confidence}
C = P_{object}\times IoU
\end{equation}
where $P_{object}$ is the probability that the box contains an object. The model considers the prediction to be a true prediction only if the box confidence score is higher than a minimum score. This minimum score is called a "threshold confidence" and is set manually. For the dataset we use here, the threshold confidence is set to $10\%\sim40\%$ to find most objects of interest while excluding unwanted objects.

The algorithm we use in this project is the state-of-the-art object detector YOLOv5 \cite{YOLOV5}, which compared to its predecessors included a new PyTorch training and deployment framework. As a result, YOLOv5 is significantly faster and user-friendly while maintaining good prediction accuracy. Therefore, YOLOv5 is regarded as one of the standard test model when developing specific algorithms in the field of fast object detection.

\section{Applications}
\begin{table*}[htbp]
  \centering
    \begin{tabularx}{0.8\textwidth} { 
  | >{\centering\arraybackslash}X 
  | >{\centering\arraybackslash}X 
  | >{\centering\arraybackslash}X
  | >{\centering\arraybackslash}X
  | >{\centering\arraybackslash}X | }
    \hline
    Labeled set size & Augmented set size & Time & Inference set of 50 accuracy & Inference set of 1000 accuracy\\
    \hline
    15 & 35 & 5 min & 68$\%$ & 52$\%$ \\
    \hline
    30 & 52 & 8 min & 98$\%$ & 85$\%$ \\
    \hline\hline
    50 & 124 & 11 min & 100$\%$ & 97$\%$ \\ 
    \hline
    120 & 299 & 15 min & 100$\%$ & 92$\%$ \\
    \hline
    200 & 449 & 28 min & 100$\%$ & 89$\%$ \\
    \hline
    \end{tabularx}%
  %
  \caption{Inference accuracy vs. dataset size. The first column reports the size of the ground-truth (manually labeled) datasets for training, validation, and test. The second column reports the size of the augmented dataset for training, validation, and test. The third column presents the run time of the training process associated with each dataset, using a Tesla T4 GPU. The last two columns report the prediction accuracy of these dataset on two inference datasets, while the inference set 1 has 50 labeled images and the inference set 2 has 1000 unlabeled images. 
  }
  \label{Benchmark}
\end{table*}%
In this section we are going to present three exemplary applications for object detection in the context of high-power laser experiments. 





\subsection{Few-cycle shadowgraphy of plasma waves}
Plasma waves excited by a laser-driven electron beam in a hybrid plasma accelerator are diagnosed. A hybrid plasma accelerator utilizes the dense, high current electron bunch produced by a laser-wakefield accelerator to drive the plasma wave for a plasma-wakefield acceleration (PWFA)\cite{hidding2019fundamentals, gilljohann2019direct, kurz2018calibration}. Unlike in PWFA driven by electron bunches from conventional RF accelerators, the plasma density in a hybrid accelerator is higher, typically $\sim10^{18}\;cm^{-3}$, which makes it possible for shadowgraphy using few-cycle optical probes\cite{savert2015direct, ding2020nonlinear, gilljohann2019direct}. The plasma evolution can be observed in detail in femtosecond resolution using a few-cycle probes beam. It is derived from the main laser driver, undergoes spectral broadening in a gas-filled fiber and is compressed to sub-10 fs by a set of chirped mirrors. Thus, probe and driver are inherently synchronized. A practical problem in experiments is the variation of the plasma waves in the shadowgrams. This especially occurs when the laser plasma parameters are being tuned, for instance, scanning the plasma target with respect to the laser focus. To locate the plasma waves regardless of the varying laser plasma condition, an object detector is used.

\subsubsection{Labeling and training}
The object detector is applied to up to 200 manually labeled shadowgrams taken from various days of experimentation. Datasets of varying sizes are used for training, and a benchmark is listed in Tab. \ref{Benchmark}. While most of the shadowgrams have observable plasma waves, about $10\%$ of the images do not. The labeled classes on the shadwgrams include the plasma waves, a shock front caused by a deliberate obstacle in the target's gas flow, and the diffraction pattern caused by dust in the imaging beam path. The dataset is split into a training set, a validation set, and a test set by $70\%$, $20\%$, and $10\%$, respectively. To further increase the size of the dataset, augmentations are applied to the labeled images, as is shown in the second column in Tab. \ref{Benchmark}. In the augmentation process, it makes copies of the original image and then slightly change the brightness and exposure. Note that augmentation is only applied to the training set but not the validation set or the test set.

The training process utilizes the concept of transfer learning, where the knowledge from a pre-trained model for general object detection tasks is transferred to our model for a specific task. YOLOv5 provides a series of such pre-trained models, and here we use the second-smallest model ("YOLO5s.pt"). The run time of the training process is listed in the third column. It is worth pointing out that the run time can be further reduced by transfer learning from a learned model using a small training set. 

\subsubsection{Results}
In addition to the test dataset, the trained models are applied to two inference datasets, as are shown in the last two columns in Tab. \ref{Benchmark}. The images in the inference sets are not used in the training, validation, or test process. The first inference set contains 50 shadwgrams with observable and labeled plasma waves. The second inference set consists of one thousand images from various experiment days, where 68 of them do not have a observable plasma wave.

\begin{figure*}[ht]
\centering
\includegraphics[width=1\linewidth]{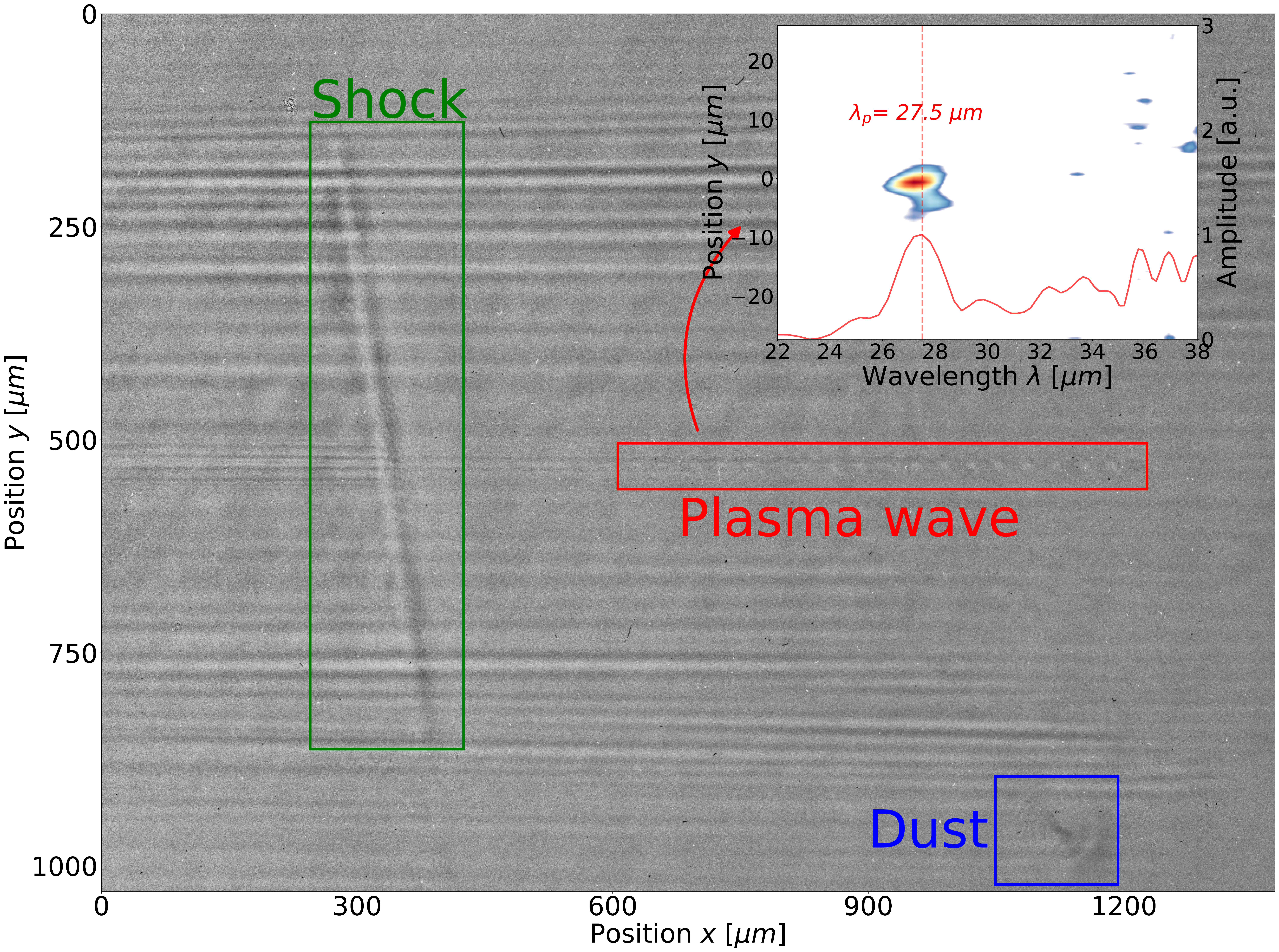}
\caption{An example: The plasma wave, the shock, and the diffraction pattern caused by dust are found by the object detector and located with bounding boxes. The subplot on the top right is the Fourier transform of the region within the bounding box of the plasma wave (red). The plasma oscillation wavelength is estimated by integrating along the vertical axis, which peaks at $27.5\mu$m. 
}
\label{plasmawave}
\end{figure*}

Comparing the five trained networks in Tab. \ref{Benchmark}, the medium-sized dataset with 50 pre-labeled shadwgrams provides the most accurate model in this case. 
The trained model has an mAP of 0.941 for an IoU threshold of 0.5. The model is then used to detect the target features on a shadowgram. An example is presented in Fig. \ref{plasmawave}, showing the detected plasma wave (red), the shock (green), and the diffraction pattern caused by dust in the beam path (blue). A threshold confidence of $10\%$ is applied when drawing the bounding boxes.


The plasma wavelength can be estimated as the plasma wave is located by the object detector. This is achieved by taking the Fourier transform of the region of interest (ROI), which is the red bounding box containing the plasma wave oscillation feature. The result of the Fourier transform is demonstrated on the top-right in Fig. \ref{plasmawave}, and the peak is at $\sim27.5\mu$m. 

\begin{figure*}[ht]
\centering
\includegraphics[width=1\linewidth]{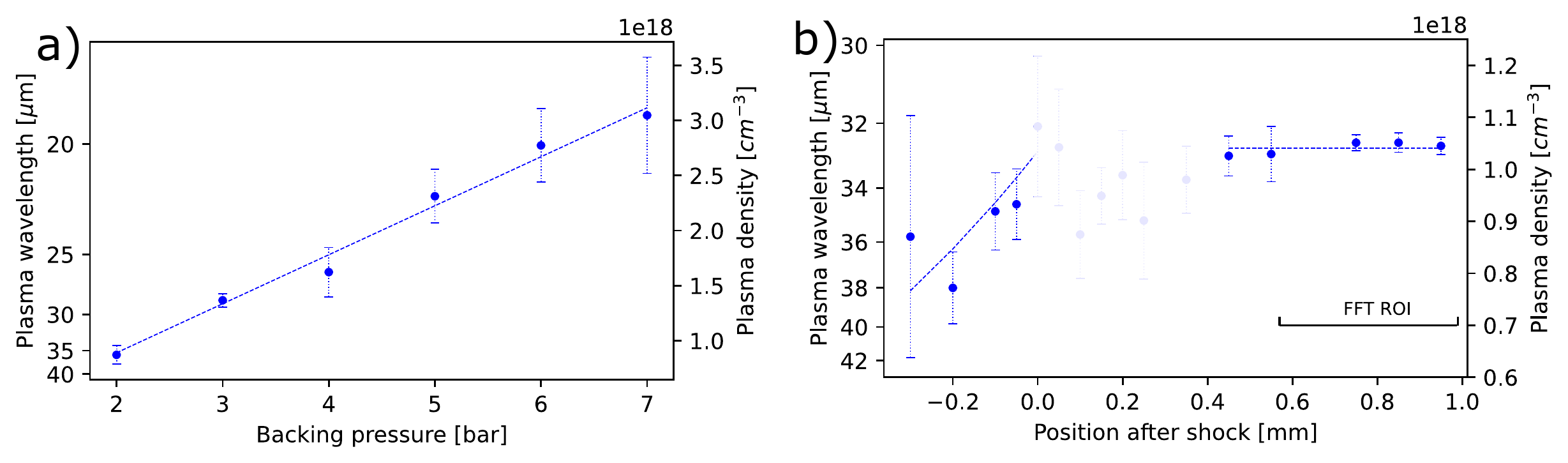}
\caption{
Plasma oscillation wavelengths (left vertical axis) and plasma density (right vertical axis), calculated from the Fourier transform results within the ROI defined by the object detector. (a) The backing pressure of the gas target is scanned from 1 bar to 6 bar. (b) The probe is moved from the upstream end to the center of the gas target, and 0 mm is where the first plasma bubble of the plasma wave is at the density shock front. As mentioned in the main text, the region where the ROI includes the shock produces unreliable results and is thus greyed out.
}
\label{ScanPositionAndDensity}
\end{figure*}

As the laser and plasma parameters (pressure, longitudinal position, etc,.) are being tuned during an experiment, the plasma wavelength changes accordingly. Fig. \ref{ScanPositionAndDensity} presents further analysis of the plasma oscillation, given the region defined by the object detector. In Fig. \ref{ScanPositionAndDensity}a, the backing pressure is scanned from 2 bar to 7 bar. Each data point represents the mean value of 20 consecutive laser shots, and the error bar measures the mean absolute deviation. The plasma wavelength is calculated by taking the Fourier transform of the plasma wave ROI at each pressure, and is plotted to the left vertical axis. The electron density is calculated from the plasma wavelength, and is labeled on the right vertical axis. Note that the right vertical axis for the plasma density profile is set to have linear tick labels, and therefore the left vertical axis for the plasma wavelength has nonlinear tick labels. The electron density vs. the backing pressure is fit to a linear relation, with a $R^2$ value as high as 0.98. The curve fitting is shown by the dashed line. 
The calculated plasma densities also agree with the interferometry measurements. 

A similar analysis is presented in Fig. \ref{ScanPositionAndDensity}b, where the few-cycle optical probe is scanned over 1.2 mm relative to the shock position, from the upstream end to the center of the gas target. At 0 mm, the first plasma bubble of the plasma wave overlaps with the density shock. The plasma density vs. position before the shock is fitted to an exponential function, and the density away from the shock approaching the target center is almost constant, both shown in blue dashed lines. The middle section of the density profile, shown as shaded circles, is lower than expected. There are two reasons for this method to be less reliable in this area. First, the shock is overlapping with the plasma wave and the width of the shock is longer than the length of a plasma bubble. Therefore, taking the Fourier transform in this area gives a wavelength longer than it should be, and thus the data points in the greyed-out region are lower than expected. Second, since the plasma wave ROI is a few hundred microns long and contains over ten plasma bubbles, the calculated plasma wavelength or density is an averaged value instead of the value at the exact probe position. A scale bar showing the length of the ROI is attached at the bottom right corner. Therefore, the peak at 0 mm near the shock is less profound than expected as it has been averaged with lower densities. At a long distance from the shock, however, the supersonic density profile is nearly constant and averaging over distance still results in a density plateau. To better resolve the density near the shock, methods that do not average over distance could be helpful, such as performing a windowed Fourier transform, a continuous wavelet transform, or even a nested object detector.

\begin{figure}[ht]
\centering
\includegraphics[width=1\linewidth]{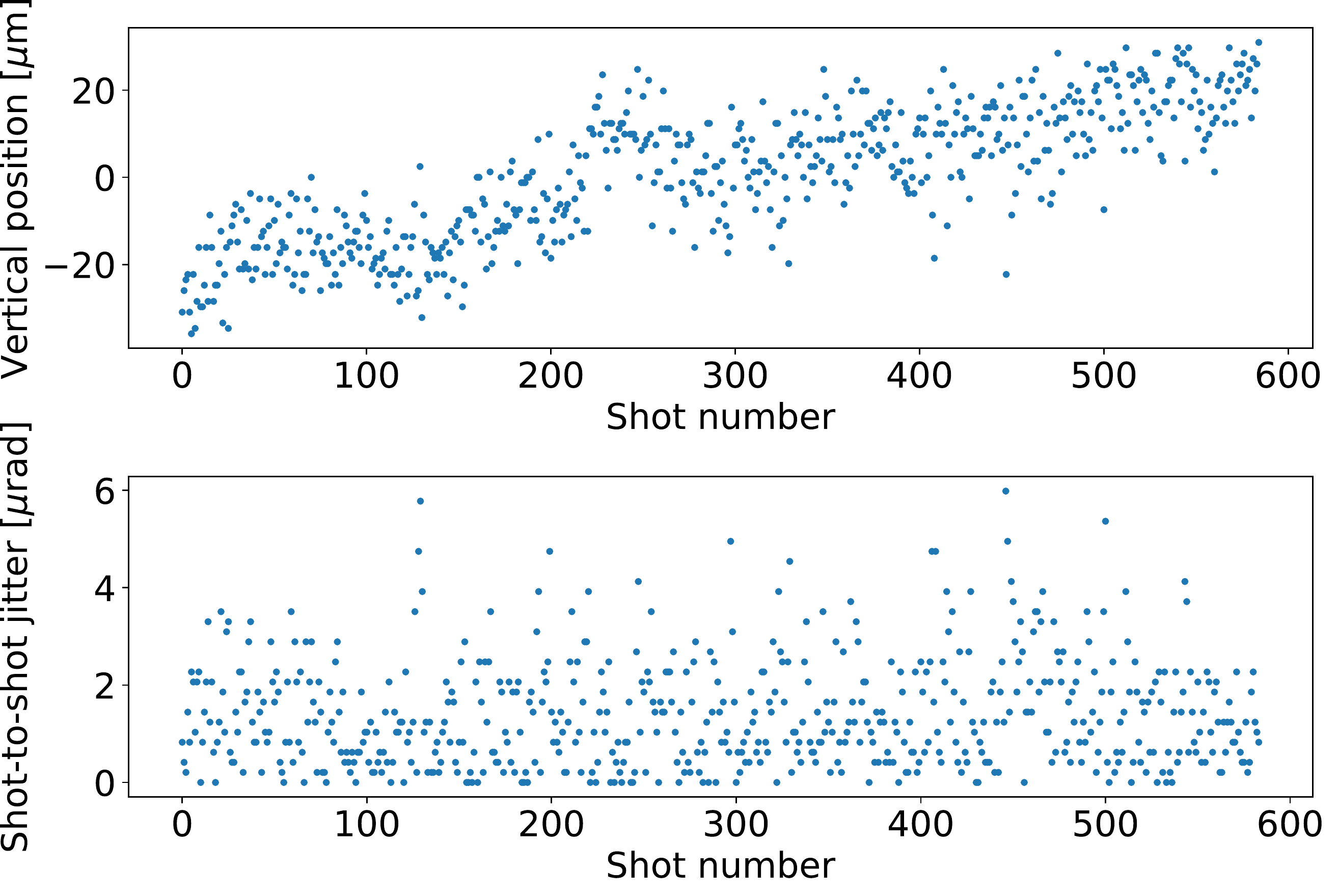}
\caption{
(a) Vertical position of the plasma wave moves over a day. (b) Jitter between every two consecutive shots, calibrated into solid angle.
}
\label{jitter}
\end{figure}
Another interesting observation is the position jitter of the plasma wave. With the object detector, the vertical position of the plasma wave can be accurately determined, leading to an estimation of the jitter of the driver. The vertical position of the plasma waves from all shots in a day is plotted in Fig. \ref{jitter}a, measured from the center of the shadowgram where a negative value means the plasma wave is on the top half of the shadowgram. Note that the parameter scan performed during the experiment only affects the horizontal position of the plasma waves, and thus there is no intentional change in the vertical position of the plasma waves. A slight slope is observed in Fig. \ref{jitter}a over the day, implying a linear drift of the plasma wave vertically. 
The focusing optic is 6 meters away from the gas jet, and the position angle is calculated accordingly. The shot-to-shot fluctuation is calculated from the difference in angle between two consecutive shots. As is shown in Fig. \ref{jitter}b, the majority of the shot-to-shot jitter falls below 4 $\mu$rad. This is in line with a separate measurement of the jitter in the laser driver, which is mostly within 3 $\mu$rad, and accordingly, the dominant source of the fluctuation in the plasma wave appears to originate from the pointing fluctuations of the main laser beam.

To summarize, the analysis in this application is enabled by the object detector, which tracks the position change of the plasma wave due to the parameter scan or even the beam jitter itself. While it is possible to select a fixed ROI for the Fourier transform without knowing the exact position of the plasma wave, such as the whole shadowgram, the estimated wavelength would not be reliable. It is due to the fact that such a maximum ROI includes too much unnecessary information, for example, the shock, the background noise, or the tail of the plasma wave without observable oscillating structures. On the other hand, the object detector locates the obvious plasma wave structures of the first few bubbles, making it possible to exclude the interference of irrelevant information during the Fourier transform. Therefore, the use of the object detector effectively increases the signal-to-noise ratio during the Fourier transform calculation.
\begin{figure*}[ht]
\centering
\includegraphics[width=1\linewidth]{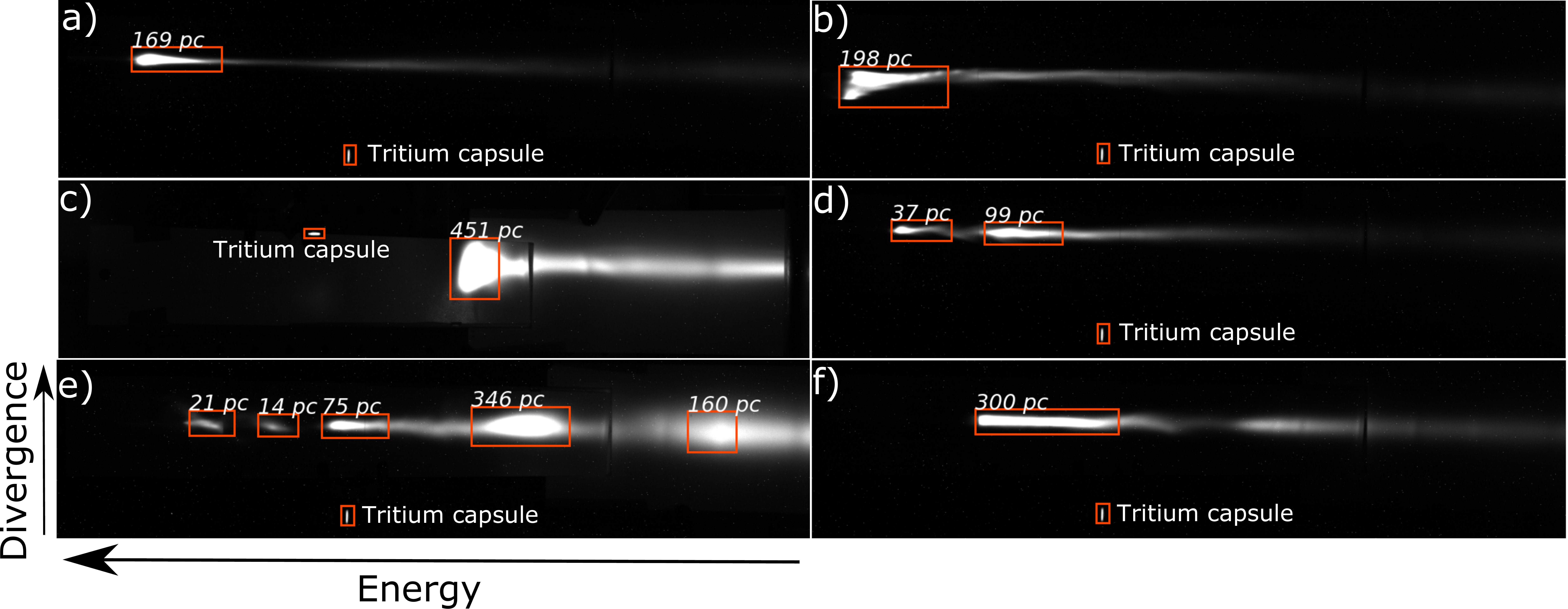}
\caption{Labeled peaks with charge number on electron energy spectra. The charge numbers are calculated from the integral within each bounding box.
}
\label{charge}
\end{figure*}
Note that the shocks and diffraction patterns caused by dust are also detected on the shadowgrams, as is shown in Fig. \ref{plasmawave}. While these objects are not analyzed in this study, they could find potential use, such as to relate the position of the shock to the position of the injection point in the accelerator\cite{wenz2019dual}. 

\subsection{Electron energy spectrometer}
The second application regards the electron beams from a laser-wakefield accelerator. The energy spectra of the produced electron beams are measured by a magnetic electron spectrometer. The magnetic spectrometer consists of an 80 cm long, 0.85 Tesla permanent magnet with 4 cm gap. The electrons get deflected as they pass through the magnetic spectrometer, and intersect with the detector plane at different position. The radius of the trajectory is determined by the energy of the electron, and thus the magnetic spectrometer is calibrated by particle tracking. 
Peaks on the electron energy spectra are identified and the associated charge numbers are calculated, not only for mono-energetic beams but also for multi-energy and broadband beams. A calibrated tritium capsule is used as a constant absolute light source in order to calibrate the detected charge. The signal of the tritium capsule is also detected and labeled on the images, from which the charge value is calculated from the pixel intensity\cite{kurz2018calibration}.


\subsubsection{Labeling and training}
The training and validation dataset consists of 50 images of electron energy spectra taken from various shot days. The two labeled classes on the energy spectra are the peaks of the electron beam and the tritium capsule. 
The dataset is expanded to 82 images using rotational augmentation.

The model is transfer-learned from the YOLO5s.pt model and then fine-tuned with a lower learning rate. Both the training and the fine-tuning process take approximately 10 minutes on a Tesla T4 GPU.
\subsubsection{Results}
The trained model has an mAP of 0.904 for an IoU threshold of 0.5. After training, the algorithm can detect peaks in the electron energy spectra, as well as the location of tritium capsule for charge calibration. The charge number of each peak is calculated and annotated alongside, see Fig. \ref{charge}. A threshold confidence of $20\%$ is applied when drawing the bounding boxes.

Fig. \ref{charge} presents six electron energy spectra with various shapes, positions, and number of peaks. The spectra were taken on different shot days. The peaks on the energy spectra are detected and labeled with the charge number, and the tritium capsule is also detected and highlighted in the small bounding box at the bottom of each image. The charge number in pico-coulomb is calibrated by the tritium capsule. Although it might be possible to estimate the charge of the peaks using pre-defined methods, applying an object detector has several advantages. First, it allows defining the region of a peak even if the peak is in an irregular shape. For instance in Fig.\ref{charge}b,f, it could be difficult to determine a peak using full-width half-max or other pre-determined definitions. Second, it enables accurate recognition of multiple peaks, such as in Fig.\ref{charge}d,e. Due to the intrinsic nature of object detection algorithms, there is no need to specify the number of peaks in advance. Last, live information on the electron charge would benefit the experimental logistics, giving extra feedback when tuning the laser-plasma parameters during experiments.

\subsection{High power laser damage on optics}
\begin{figure*}[ht]
\centering
\includegraphics[width=0.85\linewidth]{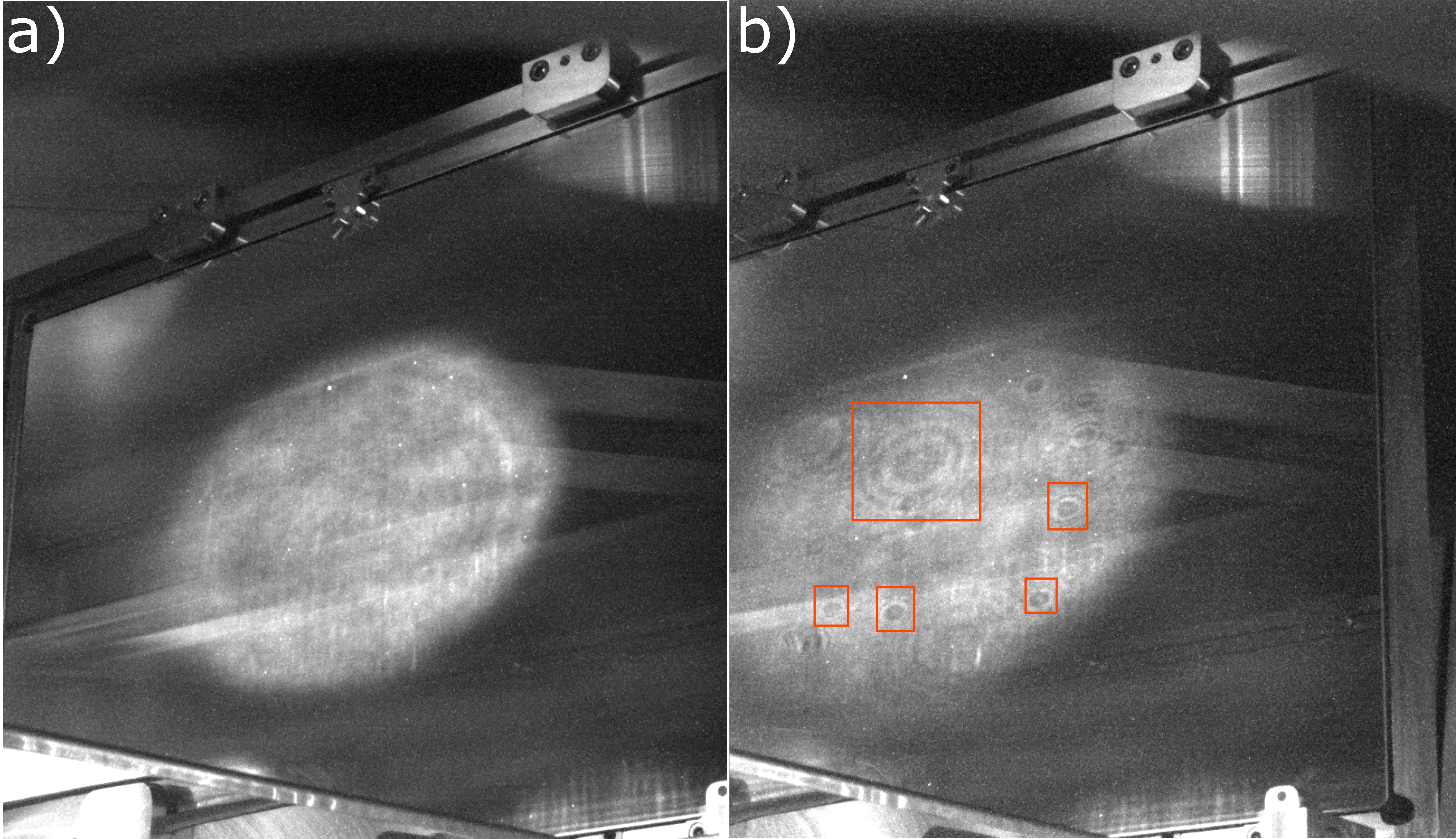}
\caption{Detected interference pattern on a grating surface, originated from damages of previous optics in the amplification beam path. 
(a) is an image of the grating surface without damaged optics in the beam path, while (b) is an image of a grating surface with damaged optics in the beam path. The bounding boxes are drawn around the detected damage spots. 
}
\label{damage}
\end{figure*}
The peak power of lasers has been ramping up since the invention of the chirped-pulse amplification technology \cite{strickland1985compression}, entering the petawatt regime in several facilities worldwide\cite{danson2004vulcan, gaul2010demonstration, sung20174, gales2018extreme, li2018339, danson2019petawatt, nees2020zeus, zhang2020laser}. 
While optics are carefully chosen for high damage threshold, the large size of PW laser optics makes optical damage a main cost driver for operating high-power laser systems. In order to minimize damage propagation along the beam path, it is therefore crucial to detect the first occurrence of a damage spot on any optic and to use such an event to trigger a laser shutdown. In this section, we present the detection of laser damage on an optic in the laser chain (wedge) by imaging the stray light off the main compressor grating and analyzing the imaging results using an object detector. In this proof-of-principle setup the damages occurred on the wedges in the amplification chain, cameras looking at the compressor grating at the end of the chain saw diffraction patterns that can be recognized by an object detector. Of course, the same imaging/object detection algorithm could be applied for directly imaging the diffuse reflection off any laser optic. 

\subsubsection{Labeling and training}
The training and validation dataset consists of 50 images of the grating surface taken from various shot days. The only labeled class on the images is the damaged spot. The dataset is expanded to 114 images using augmentation in the image-level brightness and the exposure.

The model is transfer-learned from the YOLO5s.pt model and then fine-tuned with a lower learning rate. The training takes $\sim50$ minutes and the fine-tuning takes $\sim30$ minutes on a Tesla T4 GPU.

\subsubsection{Results}
The trained model has an mAP of 0.995 for an IoU threshold of 0.5. The model is applied to two inference sets. The first inference set has 1000 images when there is no optical damage in the beam path, and the second inference set has 1000 images when there is an optical damage in the beam path. The images were taken from different shot days. The model detects no signal of damage in any of the images in the first set, while it detects the diffraction patterns from the optical damage in all 1000 images in the second set. The results prove the good consistency of the object detector since it neither misses any damaged optics nor gives false labels. Fig. \ref{damage} presents two examples of the prediction results in the inference sets, where Fig. \ref{damage}a is from the first set and Fig. \ref{damage}b is from the second set. A threshold confidence of $40\%$ is applied when drawing the bounding boxes. 
This application demonstrates the potential to use object detection in any high-power laser system for immediate warnings on crucial optical elements, providing timely protection to the rest of the optics in the system. It has to be pointed out that this application, unlike the previous two, can also be treated as a classification problem. A recent work by Tudor \textit{et al.} \cite{Tudor2022CNNLPA} finds abnormal laser beam profile using a CNN.

\section{Summary and Outlook}
In this paper, we have provided three examples of how advanced computer vision techniques can be applied to assist online and offline experiment analysis in high power laser facilities. We have shown satisfying predictions by fine-tuning pre-trained network for general object detection tasks using only $\sim50$ hand-labeled images. The learned model has been examined using not only the test dataset split from the input data, but also a separate inference set of a thousand images with various laser-plasma parameters. The model training is performed using accessible computational resources in GPU hours or below, while the inference time on an unseen image with the trained models takes only milliseconds. As we have shown, the presented methodology is widely adaptable and easy to implement. At the same time, results are reliable and robust, comparable to manual human recognition. Thus, we anticipate that the concept of object detection will find wide application in more image-related measurements and diagnostics in high power laser experiments.

\section*{Acknowledgements}
The authors would like to acknowledge support by operating resources of the Centre for Advanced Laser Applications (CALA). J. Lin would like to acknowledge support from the Alexander von Humboldt Stiftung. A. Döpp acknowledges support from the German Research Agency, DFG Project No.~453619281.

\bibliographystyle{unsrt}
\bibliography{ref.bib}


\end{document}